\documentclass[prx,aps,amsfonts,floatfix,superscriptaddress,longbibliography,twocolumn,footinbib]{revtex4-2}
\usepackage[utf8]{inputenc}
\usepackage{amsmath}    
\usepackage{amssymb}
\usepackage{mathbbol}
\usepackage{graphicx}   
\usepackage{verbatim}   
\usepackage{subfigure}  
\usepackage{hyperref} 
\usepackage{scalerel}
\usepackage{cancel}
\usepackage{soul}
\usepackage{notes2bib}
\usepackage[normalem]{ulem}
\usepackage[dvipsnames]{xcolor}
\usepackage{siunitx}
\usepackage{physics}

\usepackage{enumerate}
\usepackage{empheq}
\definecolor{blue-violet}{rgb}{0.54, 0.17, 0.89}

\begin{document}

\title{Universal spectral correlations in open Floquet systems with localized leaks}

\author{Edson M. Signor} 
\affiliation{Department of Physics, University of Connecticut, Storrs, Connecticut 06269, USA.}

\author{Miguel A. Prado Reynoso} 
\affiliation{Nonlinear Dynamics, Chaos and Complex Systems Group, Departamento de F\'{i}sica,  Universidad Rey Juan Carlos, Tulip\'{a}n s/n, 28933 M\'{o}stoles, Madrid, Spain}
\affiliation{Instituto de Ciencias Nucleares, Universidad Nacional Aut\'onoma de M\'exico, 04510 Cd. Mx., Mexico}

\author{Bidhi Vijaywargia}
\affiliation{Department of Physics, University of Connecticut, Storrs, Connecticut 06269, USA.}

\author{Sandra D. Prado}
\affiliation{Instituto de Física, Universidade Federal do Rio Grande do Sul, Porto Alegre CP15051, Brazil.}

\author{Lea F. Santos}
\affiliation{Department of Physics, University of Connecticut, Storrs, Connecticut 06269, USA.}


\begin{abstract}
We show that introducing a localized leak in Floquet systems with time-reversal symmetry leads to universal spectral correlations governed by the non-Hermitian symmetry class $\mathrm{AI}^{\dagger}$, associated with complex-symmetric Ginibre random matrices, rather than by the unconstrained Ginibre ensemble. As a concrete example, we analyze the leaky quantum standard map (L-QSM) of the kicked rotor. Since the closed map exhibits circular orthogonal ensemble (COE) statistics, the open system is naturally compared with the truncated circular orthogonal ensemble (TCOE), which models localized leakage by removing columns from a COE matrix. We find excellent agreement between the bulk spectral properties of the L-QSM and the TCOE, and demonstrate that their short-range spectral correlations follow the universal statistics of the non-Hermitian symmetry class $\mathrm{AI}^{\dagger}$. This agreement holds for smaller leak sizes as the matrices increase, while the COE limit is recovered only when the truncation is smaller than one full column. In contrast to local properties, the global density of states of the L-QSM and the TCOE approaches the Ginibre circular law only when the leakage becomes sufficiently strong. 
\end{abstract}

\maketitle


\section{Introduction}

Quantum chaos refers to properties of the spectra and eigenstates of quantum systems that reflect the chaotic dynamics of their classical counterparts.
In particular, spectral correlations that follow the predictions of random matrix theory serve as signatures of underlying classical chaos. For isolated quantum systems described by time-independent Hamiltonians, chaos implies spectral fluctuations that obey Wigner-Dyson statistics, while the correlations of the quasienergies of periodically driven (Floquet) systems are consistent with Dyson's circular ensembles~\cite{Wigner1951P,Dyson1962,MehtaBook,HaakeBook,Guhr1998}. When a quantum system is opened to an external environment, its dynamics becomes non-unitary,  the spectrum becomes complex~\cite{Grobe1988,Grobe1989,Akemann2019,Sa2020, Denisov2019, Sa2020JPA, Rubio2022, Garcia2022, Kawabata2023, Prasad2022,Wold2025,
Prosen2010, Can2019, Vallejo-Fabila2024, Villasenor2024, Roccati2024,Villasenor2024b,Mondal2025PRE,Ferrari2025,Villasenor2025,Mondal2026,Mondal2026arxiv}, 
and the eigenstates are no longer orthonormal~\cite{Brody2014,DeTomasi2022, Mak2024}, motivating extensions of quantum chaos diagnostics to the non-Hermitian domain.

According to the Grobe-Haake-Sommer conjecture \cite{Grobe1988,Grobe1989,Akemann2019,Sa2020, Denisov2019, Sa2020JPA, Rubio2022, Garcia2022, Kawabata2023, Prasad2022,Wold2025, Villasenor2025}, first introduced in the context of the periodically kicked top with damping~\cite{Grobe1988}, the spectral statistics of open quantum systems that are chaotic in the classical limit follow those of Ginibre unitary ensembles (GinUE)~\cite{Ginibre1965,Fyodorov1997PRL,Fyodorov2003}. These ensembles are characterized by a uniform distribution of the eigenvalues over the complex plane and cubic level repulsion. However, subsequent refinements and tests of this conjecture have revealed a broader picture. Ginibre-like level repulsion is not an exclusive signature of classical chaos, as it can also arise in non-Hermitian spectra without an underlying classical chaotic dynamics~\cite{Villasenor2024b,Mondal2025PRE,Ferrari2025,Villasenor2025} provided transient chaos is present~\cite{Mondal2026,Mondal2026arxiv}. In parallel, and closer in spirit to the present study, additional non-Hermitian universality classes have been identified~\cite{Bernard2002,Hamazaki2020}, giving rise to spacing distributions that differ from those of the unconstrained Ginibre ensembles~\cite{Hamazaki2020,Kanazawa2021,Akemann2022,Akemann2024,Akemann2025}.

In this work, we consider Floquet quantum systems whose spectral correlations, in the absence of dissipation, follow those of the circular orthogonal ensemble (COE). We investigate the spectral properties of these systems when they are opened. Rather than describing the coupling to an environment through master equations, we consider the less explored case of openness  introduced through a spatially localized leak, which allows probability to escape, as realized in leaky optical or microwave billiards~\cite{Altmann2013,Holmes2025,Fyodorov1997,Fyodorov1999}. Our central question is whether universal spectral correlations can still emerge in the presence of such localized leakage and, if so, which non-Hermitian universality class governs them. We find that the appropriate random-matrix benchmark is not the unconstrained complex Ginibre ensemble, corresponding to symmetry class A, but the transpose-symmetric Ginibre ensemble of class AI$^{\dagger}$.

\begin{figure*}[t]
    \centering
    \includegraphics[scale=.6]{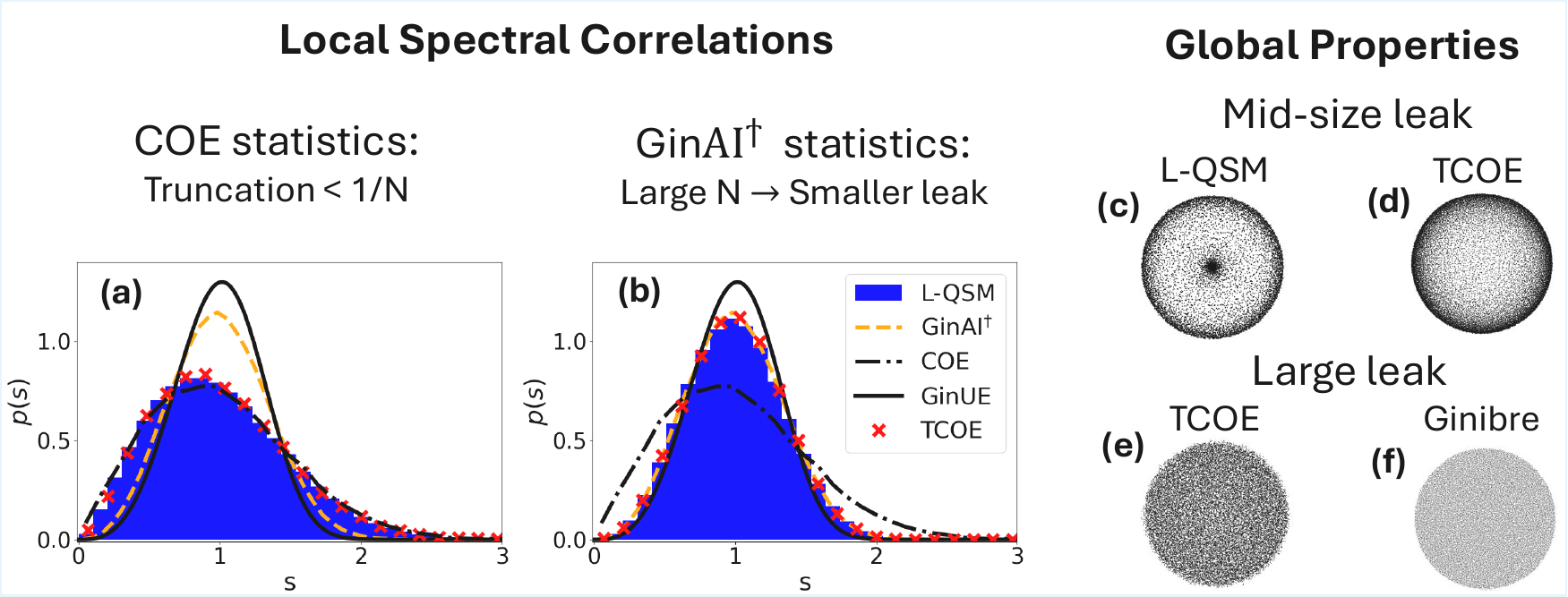}
    \caption{Sketch of the spectral properties of Floquet systems with localized leaks. 
    (a)-(b) Local spectral correlations. 
    (a) The  truncation of the unitary operator needs to be smaller than one full column for the leaky quantum standard map (L-QSM) and the truncated circular orthogonal ensemble (TCOE) to approach COE statistics. 
    (b) As the matrices dimensions grow, smaller leak sizes ensure universal spectral correlations as in $\text{GinAI}^{\dagger}$. 
    (c)-(f) Global distribution of eigenvalues. 
    (c)-(d) For intermediate leaks, the eigenvalues of the (c) L-QSM and (d) TCOE concentrate near the unit circle, unlike Ginibre matrices.
 For large leaks, (e) the eigenvalue distribution becomes uniform as in (f) Ginibre matrices.
   }
    \label{fig:Ldistri}
\end{figure*}

To illustrate these ideas, we analyze the spectrum of the strongly chaotic quantum standard map, obtained from the one-period Floquet operator of the quantum kicked rotor~\cite{Casati1979}. The system has a localized leak implemented by truncating the unitary Floquet propagator and setting to zero the columns associated with the position of the opening~\cite{Novaes2013}. The resulting operator is non-unitary and its eigenvalues correspond to complex resonances. Because the closed standard map exhibits spectral correlations consistent with the COE, the appropriate random-matrix benchmark for the open system is the truncated circular orthogonal ensemble (TCOE)~\cite{Zyczkowski2000,Gluck1999,Killip2017, Fischmann2012}, in which columns of a COE matrix are removed to model localized escape.

We compare the global statistical properties, captured by the density of states and the distribution of eigenvalues in the complex plane, and the local spectral correlations of the leaky quantum standard map (L-QSM) with those of the TCOE, the Ginibre ensemble with transpose symmetry  (GinAI$^\dagger$), and the Ginibre unitary ensemble (GinUE). A summary of the results is sketched in Fig.~\ref{fig:Ldistri} and listed below. 
\begin{itemize}
    \item 
{\bf COE statistics in Fig.~\ref{fig:Ldistri}(a):} The truncation of the unitary operator needs to be smaller than one full column for the spectral properties of the L-QSM and TCOE to approach those of the closed system, recovering COE statistics.

\item {\bf GinAI$^{\dagger}$ statistics in Fig.~\ref{fig:Ldistri}(b):} The short-range correlations in the bulk of both the L-QSM and TCOE spectra show excellent agreement with the symmetry-constrained GinAI$^{\dagger}$ statistics and differ from the unconstrained GinUE. The leak size required to reach this agreement decreases as the matrix dimension increases.

\item {\bf Non-uniform distributions in Figs.~\ref{fig:Ldistri}(c)-(d):}
For intermediate truncation strengths, the eigenvalues of the L-QSM and TCOE show concentration of long-lived resonances near
the unit circle, which contrasts with that of Ginibre ensembles, where the eigenvalues are uniformly distributed over the complex plane, according to the circular law. 

\item {\bf Uniform distributions in Fig.~\ref{fig:Ldistri}(e)-(f):} Motivated by the analysis in~\cite{Zyczkowski2000}, we investigate the spectral properties of the TCOE when a large fraction of columns of the underlying COE matrix are truncated. In this case, the density of states approaches the Ginibre circular law, which is independent of the symmetry class. The local spectral correlations
\end{itemize}

The paper is organized as follows. Section~\ref{Sec:Map} reviews the classical and quantum standard map and introduces the leaky version of the model. Section~\ref{Sec:statisticsQSM} compares the density of states and spectral correlations of the L-QSM with those of the TCOE and with Ginibre ensembles in the non-Hermitian symmetry classes A and AI$^\dagger$. Section~\ref{Sec:COEcrossover} analyzes how the short-range spectral correlations of the L-QSM and TCOE depend on the leak size and when COE statistics is recovered. Section~\ref{Sec:LargeLeakage} examines the approach of the TCOE density of states toward the Ginibre form as the leakage increases. 
Our conclusions are presented in Sec.~\ref{Sec:Conclusions}.

\section{Classical and quantum standard map with leakage}
\label{Sec:Map}

We begin by briefly reviewing the closed standard map in the classical and quantum domains, followed by a description of the system when a leak is introduced. More details can be found in Appendix~\ref{App:QSM}.

\subsection{Closed standard map}

A classical kicked rotor consists of a rotor that receives periodic instantaneous ``kicks'' and moves freely between them. Its Hamiltonian is given by
\begin{equation}
    H(q,p) = \frac{p^2}{2} - \frac{K}{4\pi^2}\cos{(2\pi q)}\sum_{n = -\infty}^{\infty}\delta(t - n) ,\label{H_cl}
\end{equation}
where $q$ is the angular position, $p$ is the conjugate momentum, $K$ is the kick strength that controls the nonlinearity of the system, and the kicking period is set to unity. Integrating Hamilton's equations over one period, from just after kick $n$ to just after kick $n+1$, we can construct the discrete-time standard map (Chirikov map)~\cite{Chirikov1971,Reichl2004},
\begin{equation}
\begin{split}
q_{n+1} & =  q_n + p_{n}  \quad \pmod 1
\\
p_{n+1} & = p_n -  \frac{K}{2\pi}\sin{(2\pi q_{n+1})}  \quad \pmod 1 ,
\end{split}
\label{eq:SM}
\end{equation}
where both canonical variables $(q,p)$ are taken modulo $1$, confining the motion to a toroidal phase space $\mathbb{T}^2$ with dimensionless coordinates $q,p \in [0,1)$. 

The parameter $K$ governs the transition from regular to chaotic motion. For $K = 0$, the dynamics is integrable. As $K$ increases, the system enters a mixed phase-space regime, characterized by the coexistence of chaotic seas and regular islands. For sufficiently large $K$, the stable islands are destroyed and the dynamics become fully chaotic. In this work, we focus on the strongly chaotic regime $K=10$.

The quantum kicked rotor is obtained by quantizing the classical Hamiltonian in Eq.~(\ref{H_cl}). Because the classical phase space of the standard map is compact with unit area, quantization restricts the system to a finite Hilbert space of dimension $N=1/(2\pi\hbar)$, where each one of the $N$ quantum states occupies a minimal phase-space area of size $2\pi \hbar$ with $\hbar$ being the effective Planck constant~\footnote{The effective $\hbar$ incorporates the physical scales of the rotor (period and moment of inertia), which are set to unity in our dimensionless formulation of the standard map in Eq.~(\ref{eq:SM})}. 

The canonical commutation relation $[\hat{q},\hat{p}]=i\hbar$ is enforced, and the toroidal phase space leads to periodic boundary conditions in both position and momentum. As a result, both variables become discrete, taking values on the uniform grid,
\[
q_n = \frac{n+\varphi}{N}, \quad \quad p_m = \frac{m}{N}, \quad \quad n,m = 1, \dots, N,
\]
where $\varphi$ specifies the boundary-condition phase associated with torus quantization. Specifically, the quantum standard map requires the wavefunctions to satisfy $\psi (q+1) = e^{2\pi i \varphi} \psi (q)$. In our calculations, we fix $\varphi =0.25$ to avoid symmetries associated with periodic boundary conditions ($\varphi =0$)~\cite{Casati1979,Berry1979}.

The system evolves under the Floquet operator of the quantum standard map, which for a unit period takes the form (see Appendix~\ref{App:QSM})
\begin{equation}
U_{n,n'} \!= 
\sqrt{\frac{i}{N}} 
\exp \!\left\{ \!
    \frac{i\pi}{N}(n-n')^2 
    + \frac{iNK}{2\pi} 
    \cos \! \left[ \frac{2\pi}{N}(n+\varphi) \right] \!\right\} .
\label{eq:QSM_closed}
\end{equation}
The spectral decomposition of the Floquet operator, 
\begin{equation}
\hat{U}|\phi_\mu\rangle =  e^{i\vartheta_{\mu}}|\phi_\mu\rangle,
\end{equation}
yields the quasienergies (or eigenangles) $\vartheta_{\mu} \in [-\pi , \pi)$ and the corresponding Floquet eigenstates $|\phi_\mu\rangle$. 
The phase-space structures of these eigenstates exhibit a clear correspondence with the underlying classical dynamics (see Appendix~\ref{App:QSM}). 

\subsection{Open standard map with a single leak}

\begin{figure}[h]
    \centering
    \includegraphics[scale=.4]{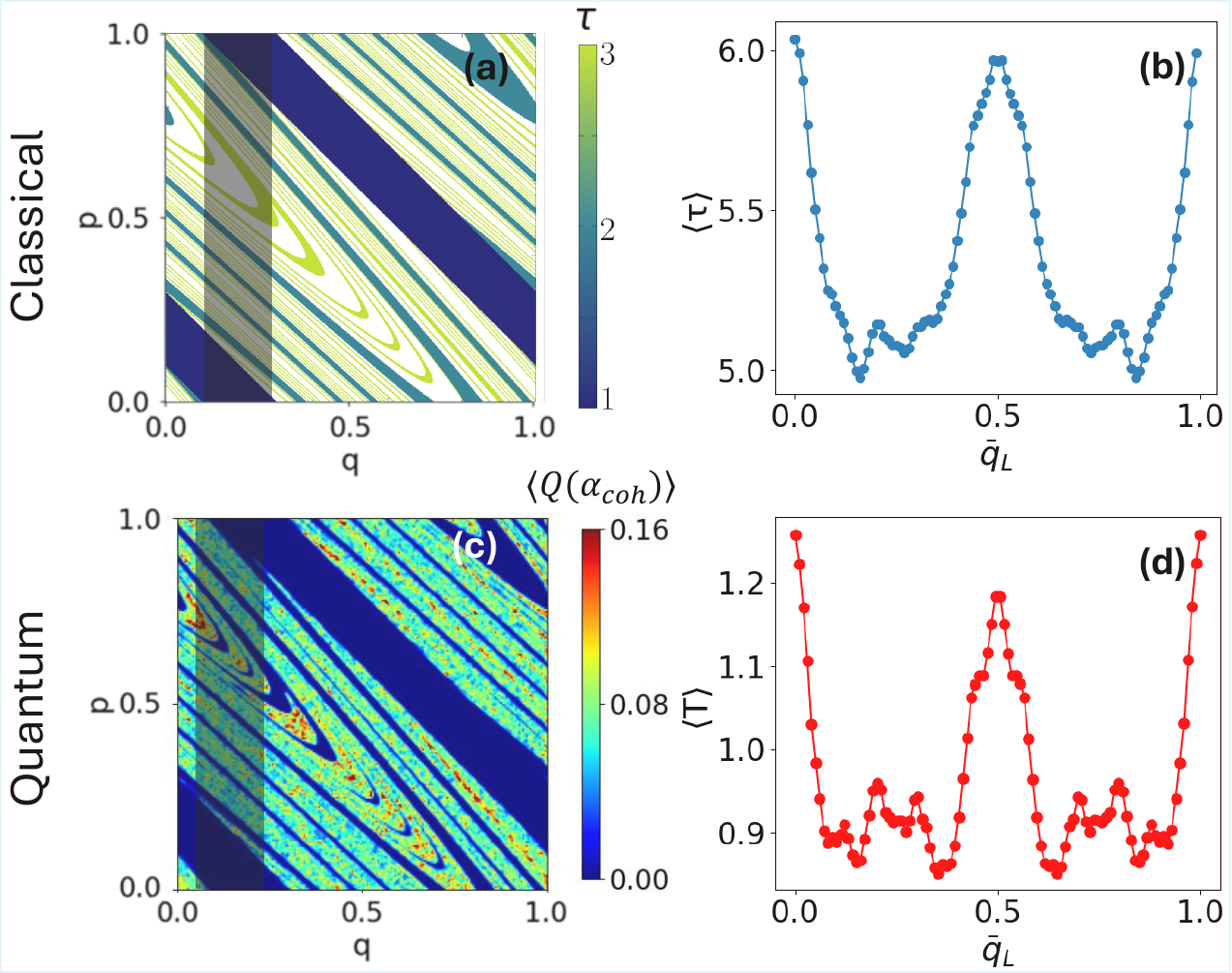}
    \caption{Open (a)-(b) classical and (c)-(d) quantum standard map with leakage; $\Delta q=0.2$, $K=10$, $N=10^4$. (a) Map of the classical dwell time $\tau$ for a slit (shaded area) centered at $\bar{q}_L = 0.2$. Dark blue indicates trajectories that escape after $\tau=1$ iteration; blue for $\tau=2$; green for $\tau=3$; and white corresponds to long-lived trajectories ($\tau > 3$). (b) Classical dwell time averaged over the phase space as a function of the slit position $\bar{q}_L$. (c) Husimi Q-function averaged over the 20 longest-lived eigenstates. (d) Quantum dwell time averaged over all eigenstates as a function of the slit position $\bar{q}_L$. 
    }
    \label{fig:OpenSM}
\end{figure}

We open the system by introducing a leak of width $\Delta q$ centered at position $\bar{q}_{L}$. In phase space, this leak corresponds to a vertical strip parallel to the momentum axis, as shown with a shaded rectangle centered at $\bar{q}_L=0.2$ in Figs.~\ref{fig:OpenSM}(a,c). Classically, the dynamics follow those of the closed map until a trajectory enters the leak region, $q \in [\bar{q}_{L} - \Delta q/2, \bar{q}_{L} + \Delta q/2]$, at which point the trajectory is terminated, representing the escape of the particle through the opening~\cite{Altmann2013,Schomerus2004}.

The quantum counterpart of the classical open standard map is described by a non-unitary Floquet operator denoted by $\widetilde{U}$. The opening is modeled by a projection operator $\hat{\Pi}$ that eliminates probability amplitude in the leaked region of phase space~\cite{Novaes2013}. The resulting open-system propagator is therefore given by
\begin{equation}
    \widetilde{U} = \hat{U} \hat{\Pi}.
    \label{open_p}
\end{equation}
Since the leak corresponds to a vertical strip parallel to the momentum axis in phase space, the projection operator is diagonal in the position basis, and its diagonal elements are
\[
\Pi_{nn} =
\begin{cases}
0, & \text{if } n \in \text{leak region},\\[4pt]
1, & \text{otherwise}.
\end{cases}
\]
Consequently, $\widetilde{U}$ coincides with the closed-system propagator except for the columns corresponding to the leak positions which are set to zero. These positions are given by the integer indices
\begin{equation}
n = N q \in 
\left\lfloor
    N\!\left(\bar{q}_{L} - \frac{\Delta q}{2},\, 
    \bar{q}_{L} + \frac{\Delta q}{2}\right)
\right\rfloor,
\end{equation}
where $\lfloor x \rfloor$ indicates the integer part of $x$. We refer to each column of zeroes as an open channel.

In the spectrum of the operator $\widetilde{U}$,
\begin{equation}
    \widetilde{U}|\phi_\mu^{R} \rangle = \lambda_\mu |\phi_\mu^{R}\rangle = e^{i\varepsilon_\mu} |\phi_\mu^{R}\rangle,
\end{equation}
the complex Floquet eigenvalues (Floquet resonances) $\lambda_\mu$ no longer lie on the unit circle, leading to $|\lambda_\mu|\leq1$, as evident in Fig.~\ref{fig:Ldistri}(a2). The quasienergies $\varepsilon_\mu$ are now complex,
\begin{equation}
\varepsilon_\mu = \vartheta_\mu + \frac{i}{2}\Gamma_\mu,
\label{eq:gamma}
\end{equation}
where $\vartheta_\mu \in [-\pi,\pi)$ and $\Gamma_\mu > 0$ gives the decay rate of each resonance, 
defining the quantum dwell time 
\begin{equation}
T_\mu = \frac{1}{\Gamma_\mu } .  
\label{eq:dwelltime}
\end{equation}
The right and left eigenstates of $\widetilde{U}$, denoted 
$|\phi_\mu^{R}\rangle$ and $|\phi_\mu^{L}\rangle$, form a biorthogonal set, 
where $|\phi_\mu^{R}\rangle$ describes forward evolution and 
$|\phi_\mu^{L}\rangle$ corresponds to backward propagation~\cite{Schomerus2004, Hall2023}.

\subsubsection{Stickiness}

Figure~\ref{fig:OpenSM} shows that structures in the phase space that are barely perceptible in the closed system due to rapid mixing are revealed once a leak is introduced. As shown in Ref.~\cite{Prado2024}, even in the strongly chaotic regime, the standard map exhibits stickiness, that is, inhomogeneous regions in phase space, where chaotic trajectories can become temporarily trapped before eventually escaping. In closed systems, these structures are hidden by the fast mixing of trajectories, but when a leak is introduced, the trajectories are terminated as soon as they reach the leak, interrupting the mixing process. This effectively ``freezes'' the short-time imprint of the phase-space inhomogeneities and makes the sticky structures visible. Stickiness affects both classical and quantum dynamics.
 
The effect of stickiness to the classical dynamics is illustrated in Figs.~\ref{fig:OpenSM}(a)-(b). Figure~\ref{fig:OpenSM}(a) shows the classical dwell time for trajectories initialized across the phase space. The escape times vary significantly. Dark blue corresponds to trajectories that escape after a single iteration ($\tau=1$); lighter blue to those escaping after two iterations ($\tau=2$); green for three ($\tau=3$); and white indicates long-lived trajectories ($\tau > 3$). This spatial variation reveals the inhomogeneity of the chaotic phase space.
The same behavior is quantified in Fig.~\ref{fig:OpenSM}(b), which shows the average classical dwell time as a function of the leak position. The dwell time is systematically larger when the leak covers regions of phase space associated with enhanced stickiness.

The quantum dynamics mirrors the classical behavior. In Fig.~\ref{fig:OpenSM}(c), we show the Husimi $Q$-function 
\begin{equation}
Q_{\mu}(\alpha_{\text{coh}}) =  \frac{N}{2\pi}\, |\langle \alpha_{\text{coh}}|\phi_{\mu} \rangle|^2,
\end{equation}
where $|\alpha_{\text{coh}}\rangle$ denotes a coherent state centered at the phase-space point $(q,p)$ \cite{Husimi1940,Zyczkowski1986, Saraceno1990, Backer2002}. The average of the Husimi distributions over long-lived resonances displays enhanced probability density (red dots) in the same regions where the classical dwell time is largest.  The quantum-classical correspondence is further demonstrated in   Fig.~\ref{fig:OpenSM}(d), where the average quantum dwell time exhibits the same dependence on the position of the leak as the average classical dwell time in Fig.~\ref{fig:OpenSM}(b). The leak therefore acts as a probe that exposes the fine structure of the classical chaotic phase space, revealing the influence of stickiness on both classical and quantum dynamics.

Our initial motivation for the present work was to investigate whether stickiness can influence spectral properties. As discussed later, we find that it has an effect on global spectral features, particularly on the distribution of eigenvalues in the complex plane. However, stickiness does not significantly affect local spectral statistics, such as short-range eigenvalue correlations.

\section{Spectral properties of the leaky quantum standar map}
\label{Sec:statisticsQSM}

In this section, we analyze the complex spectrum of the leaky quantum standard map (L-QSM) and compare it with random-matrix benchmarks. Since our interest is on the universal properties of the bulk of the spectrum, very short-lived resonances are discarded, as explained in Appendix~\ref{sec:bulk_selection}.

\subsection{Random matrix comparison}
\label{subsec:rmt_benchmarks}
To characterize the spectral properties of the L-QSM, we compare its density of states and spectral correlations with those of the truncated circular orthogonal ensemble (TCOE),  the Ginibre unitary ensemble (GinUE), and the Ginibre ensemble with transpose symmetry  (GinAI$^\dagger$).  

\subsubsection{Truncated circular orthogonal ensemble (TCOE)}
Because the standard map originates from a time-reversal invariant Hamiltonian, the spectrum of its unitary Floquet propagator $\hat{U}$ is expected to follow the circular orthogonal ensemble (COE) statistics. Therefore, when a localized leak is introduced, the corresponding non-unitary operator of the L-QSM, $\widetilde{U}$, is naturally compared with the TCOE. In this ensemble, $N \times N$ matrices are drawn from the COE, and $n_L$ columns have their elements set to zero. The truncation ratio $n_L/N$ directly corresponds to the leak size $\Delta q$ in the open map,
\begin{equation}
    \Delta q = \frac{n_L}{N}.
\end{equation}

\subsubsection{Ginibre unitary ensemble (GinUE)}

The unconstrained complex Ginibre ensemble GinUE, associated
with non-Hermitian symmetry class $A$, consists of $N\times N$ matrices
whose entries are independent complex Gaussian random variables~\cite{Ginibre1965,Khoruzhenko2009},
\begin{equation}
G_{ij}=\frac{\mathcal{N}(0,1)+i\,\mathcal{N}(0,1)}{\sqrt{2N}},
\label{eq:GinibreA}
\end{equation}
where \(\mathcal{N}(0,1)\) denotes a real Gaussian random variable with zero
mean and unit variance. Thus, the real and imaginary parts of each matrix
element are independent and have variance \(1/(2N)\), so that
\(\langle |G_{ij}|^2\rangle = 1/N\).

\subsubsection{\texorpdfstring{Ginibre ensemble with transpose symmetry (GinAI$^\dagger\!$)}{Ginibre ensemble with transpose symmetry (GinAIdagger)}}

A complex Ginibre matrix in the non-Hermitian symmetry class
$\mathrm{AI}^\dagger$ is obtained by imposing the transpose symmetry
\begin{equation}
G_{\mathrm{AI}^\dagger}^{\,T}=G_{\mathrm{AI}^\dagger}.
\label{eq:AIdagger_symmetry}
\end{equation}
This condition means that the matrix is complex symmetric. 

One convenient way to generate such matrices is to start from an unconstrained
complex Ginibre matrix \(X\) drawn from class \(A\) with the same variance
convention as in Eq.~\eqref{eq:GinibreA}, and then symmetrize it,
\begin{equation}
G_{\mathrm{AI}^\dagger}=\frac{X+X^T}{\sqrt{2}}.
\label{eq:AIdagger_construction}
\end{equation}
Equivalently, one may generate the upper-triangular part of the matrix using
independent complex Gaussian variables and then copy it to the lower-triangular
part. The diagonal elements remain complex Gaussian variables, while the
off-diagonal elements satisfy \(G_{ij}=G_{ji}\).

\subsection{Density of states}
\label{sec_DOS}

In the large-$N$ limit, the density of states
\begin{equation}
    R_{1}(\lambda) = \sum_{\mu} \delta(\lambda - \lambda_\mu)
\end{equation}
of both GinUE and GinAI$^\dagger$ lead to the Ginibre circular law,
\begin{equation}
    R_1(\lambda)=\frac{1}{\pi}.
\end{equation}
This is because the eigenvalues of Ginibre random matrices, independently of their symmetry class, are uniformly distributed in the complex plane. 

Rather than visualizing the full two-dimensional density of states in a three-dimensional plot, we compute $R_1(\lambda)$ from the complex eigenvalues and plot it as a function of their modulus $|\lambda|$. Since different eigenvalues may have the same absolute value, multiple points can fall into the same $|\lambda|$ bin. We use bins of width $10^{-2}$. 

\begin{figure}[h]
    \centering
    \includegraphics[scale=.33]{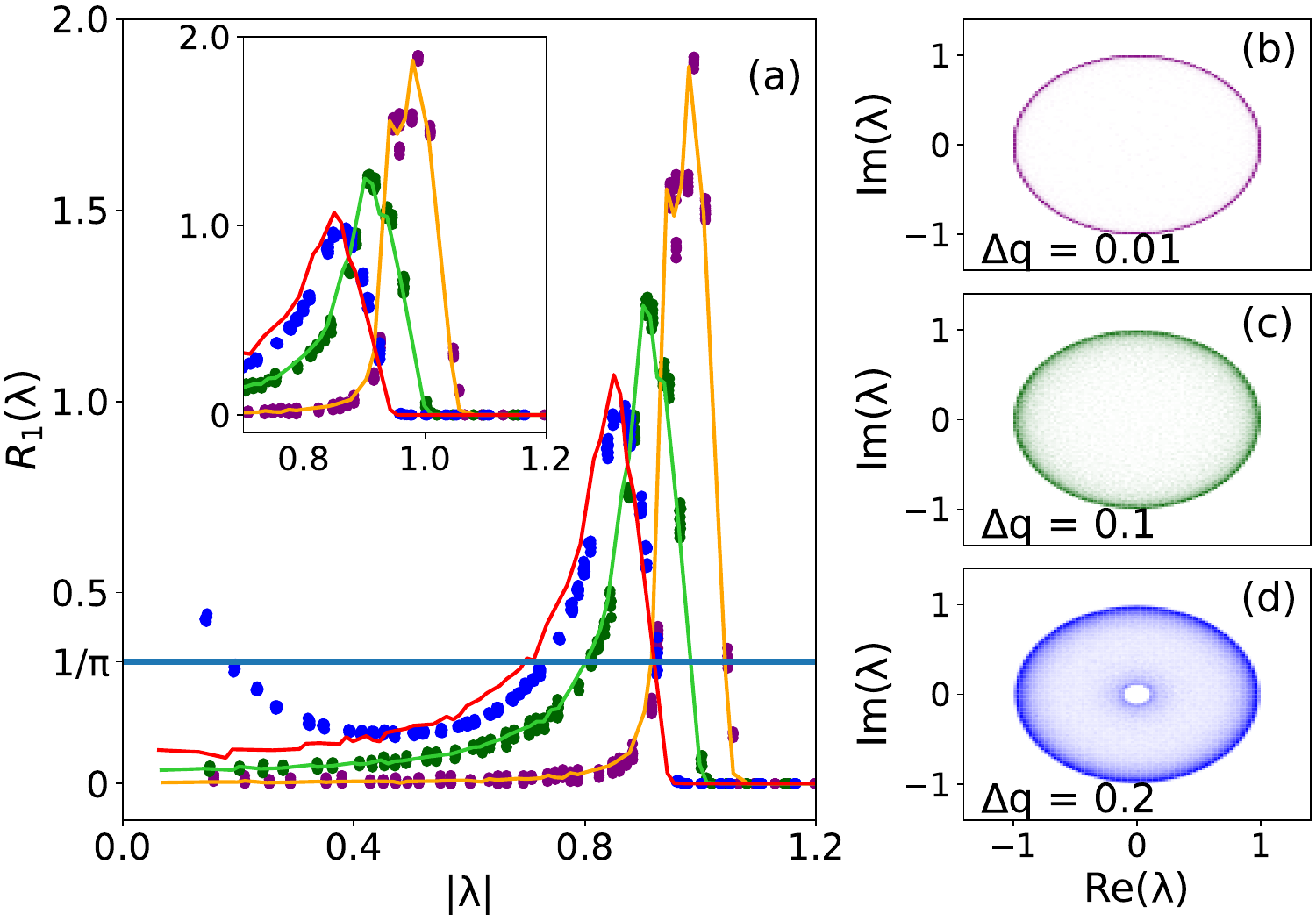}
    \caption{(a) Averaged density of states $R_1 (\lambda)$ of the complex eigenvalues $\lambda$ of the L-QSM (circles), TCOE (solid curves), and GinUE (horizontal line at $1/\pi$) plotted as a function of the modulus $|\lambda|$. The leak is centered at $\bar{q}_L = 0.2$ and three slit sizes are shown: $\Delta q = 0.2$ (blue circles and red line), $\Delta q = 0.1$ (green circles and light-green line), and $\Delta q = 0.01$ (purple circles and orange line). The agreement between the L-QSM and the TCOE progressively extends to the region of low $|\lambda|$ as $\Delta q$ decreases. The inset highlights the region of long-lived resonances, where there is excellent agreement between QSM and TCOE for all slit sizes. (b)-(d): Rescaled two-dimensional density of states of the L-QSM for increasing slit size from (b) to (d). The white hole around the origin, clearly visible in (d), corresponds to the excluded short-lived resonances, as described in the text. For each value of $\Delta q$, multiple ensemble realizations are considered to ensure a total of at least $10^5$ eigenvalues. The QSM ensemble is generated with  $9.99 < K < 10.01$. 
    }
    \label{fig:DOS_STM_COE}
\end{figure}

Figure~\ref{fig:DOS_STM_COE}(a) compares the density of states of the L-QSM with the TCOE. The Ginibre circular law is also shown as a reference line at $1/\pi$. The circles correspond to the L-QSM data and the solid curves represent the TCOE results obtained by averaging over all points that fall into each bin.

For the slit of size $\Delta q = 0.2$ (blue circles and red curve),
Fig.~\ref{fig:DOS_STM_COE}(a) shows that:
\begin{enumerate}
    \item $R_1(\lambda)$ for the L-QSM and TCOE deviate from the uniform Ginibre distribution. Their concentration of long-lived resonances illustrated in Fig.~\ref{fig:Ldistri}(a2) and Fig.~\ref{fig:DOS_STM_COE}(d) appears as a peak in Fig.~\ref{fig:DOS_STM_COE}(a). 
    \item $R_1(\lambda)$ for the L-QSM and TCOE strongly agrees for slow-decaying resonances, that is, eigenvalues with large $|\lambda|$. In the L-QSM, these long-lived resonances correspond to classical trajectories that avoid the leak for many iterations, and therefore explore the phase space nearly ergodically. Because they remain effectively uncorrelated with the leak, their spectral properties reproduce those of the TCOE. 
    \item The agreement between L-QSM and TCOE gradually deteriorates as $|\lambda|$ decreases. 
    The short-lived resonances of the L-QSM retain some degree of correlation with the leak, leading to deviations from TCOE statistics. 
\end{enumerate}

As the size of the leak is reduced in Fig.~\ref{fig:DOS_STM_COE}(a), the agreement between the density of states of the L-QSM and the TCOE  progressively extends into the region of lower $|\lambda|$. This trend is clearly visible when comparing the results for $\Delta q=0.2$ (blue circles and red curve) with those for $\Delta q=0.1$ (green circles and light-green curve) and $\Delta q=0.01$ (purple circles and orange curve). As $\Delta q$ increases, the peak of the spectral distribution shifts toward $|\lambda| \rightarrow 1$, becoming more pronounced near the unit circle, while the density decreases at small $|\lambda|$. 

This shift reflects the gradual approach to the closed system, as illustrated by the two-dimensional density of states of the L-QSM shown in Figs.~\ref{fig:DOS_STM_COE}(b)-(d). As the leak decreases from the bottom panel (d) to the top panel (b), the eigenvalues migrate away from the origin and further accumulate near the unit circle. In the limit $\Delta q \to 0$, the effect of leakage becomes negligible and the spectrum approaches that of the closed system recovering COE statistics. This approach is further discussed in Sec.~\ref{Sec:COEcrossover}.

\subsubsection{Effect of stickiness}

All results in Fig.~\ref{fig:DOS_STM_COE} correspond to a leak centered at $\bar{q}_L = 0.2$. If the leak is instead placed at $\bar{q}_L = 0.5$ (data not shown), the peak of the density of states shifts toward larger values of $|\lambda|$, indicating a higher proportion of long-lived resonances. This behavior is consistent with the dependence of the quantum dwell time on $\bar{q}_L$ presented in Fig.~\ref{fig:OpenSM}(d), where the influence of stickiness is evident. As demonstrated in Ref.~\cite{Prado2024}, the region around $\bar{q}_L = 0.2$ exhibits stronger sticky structures than the region near $\bar{q}_L = 0.5$. When the leak covers a sticky region (as at $\bar{q}_L = 0.2$), classical trajectories are more efficiently removed, resulting in a larger fraction of short-lived resonances. Conversely, placing the leak in a less sticky region (such as $\bar{q}_L = 0.5$) preserves more long-lived resonances, shifting the spectral weight closer to the unit circle. 

\subsection{Short-range spectral correlations}
\label{sec:Spc_d}

Two commonly adopted strategies to characterize short-range spectral correlations are the nearest-neighbor eigenvalue spacing distribution and the ratio of consecutive eigenvalue spacings. Both are used here.

\subsubsection{Nearest-neighbor eigenvalue spacing distribution}
Since the eigenvalues $\lambda_\mu $ of the non-Hermitian matrices studied here lie in the complex plane, the nearest-neighbor eigenvalue spacing is defined as the minimal Euclidean distance, 
\begin{equation}
    s_\mu = |\lambda_\mu - \lambda_\mu^{\mathrm{NN}}|,
\end{equation}
between $\lambda_\mu$ and the eigenvalue $\lambda_\mu^{\mathrm{NN}}$ that is its nearest neighbor in the complex plane. As in Hermitian random-matrix theory, unfolding is necessary to separate system-specific 
spectral features from universal fluctuations, as described in Appendix~\ref{App:Unfolding}.

The nearest-neighbor eigenvalue spacing distribution of Ginibre matrices depend on their symmetry class~\cite{Hamazaki2020}. The distribution for the GinUE is given by
\begin{equation}
    \mathrm{p}_{\text{GinUE}}(s) = C \, \tilde{\mathrm{p}}(C s),
    \label{eq:GinUE}
\end{equation}
with kernel 
\begin{equation}
    \tilde{\mathrm{p}}(s) = \sum_{j=1}^{\infty} \frac{2 s^{2j + 1} e^{-s^2}}{\Gamma(j + 1, s^2)} \prod_{j=1}^{\infty} \frac{\Gamma(j + 1, s^2)}{\Gamma(j + 1)},
\end{equation}
where $\Gamma(x,y)$ is the incomplete gamma function and the constant 
\begin{equation}
    C = \int_0^{\infty} s \, \tilde{\mathrm{p}}(s)\, ds \approx 1.1429\dots
\end{equation}
ensures normalization and unit mean spacing, $\langle s \rangle = 1$. 

\begin{figure}[h]
    \centering
    \includegraphics[scale=.4]{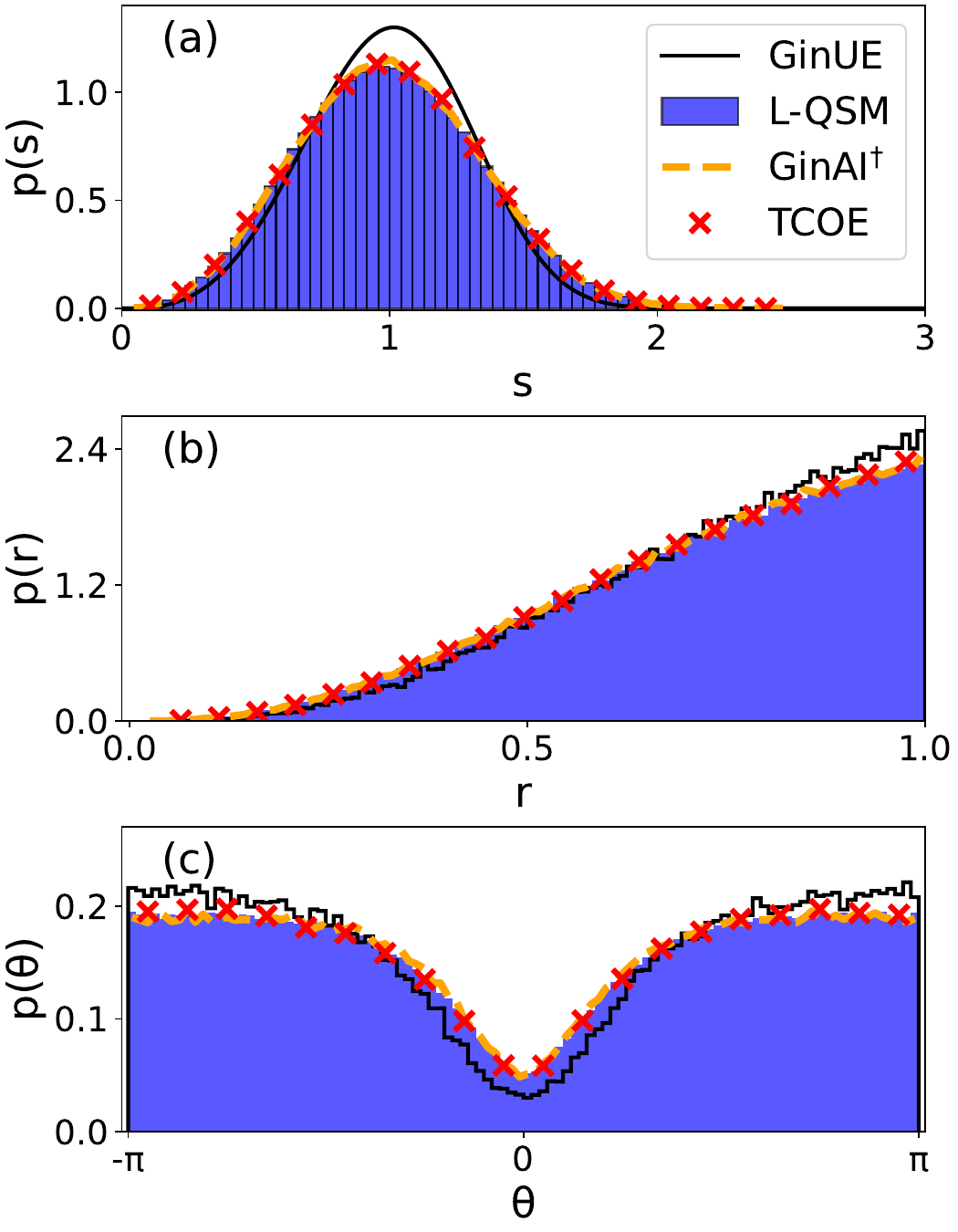}
    \caption{Short-range spectral correlations for the L-QSM with slit size and position $\Delta q = \bar{q}_L = 0.2$ (blue shade), compared with TCOE (red cross), GinUE (black solid line), and $\text{GinAI}^{\dagger}$ (orange dashed line). (a) Nearest-neighbor eigenvalue spacing; (b) radius distribution of the ratio of consecutive spacings; and (c) angle distribution of the ratio of consecutive spacings. Excellent agreement for L-QSM, TCOE, and $\text{GinAI}^{\dagger}$, while they deviate from GinUE.
     Multiple ensemble realizations are used to ensure a total of at least $10^5$ eigenvalues. The L-QSM ensemble is generated with $9.99 < K < 10.01$ and $N = 10^4$.}
    \label{fig:lvlcorr_STM}
\end{figure}

There are several symmetry classes that exhibit local statistics as in the GinUE~\cite{Hamazaki2020}. However, for the GinAI$^\dagger$, where the Ginibre matrices obey transposition symmetry, the nearest-neighbor spacing distribution $\mathrm{p}_{\text{AI}^\dagger}(s)$ deviates from the GinUE form $\mathrm{p}_{\text{GinUE}}(s)$.  Numerical simulations of large random matrices show that $\mathrm{p}_{\text{AI}^\dagger}(s)$  has a lower peak and a broader variance than the GinUE distribution, reflecting the reduced number of effective complex degrees of freedom governing eigenvalue repulsion. 

Figure~\ref{fig:lvlcorr_STM}(a) compares the nearest-neighbor eigenvalue spacing distribution of the L-QSM (blue shade) to those of the TCOE (red crosses), $\text{GinAI}^\dagger$ (orange dashed line), and
GinUE (black solid line). The distribution for the L-QSM exhibits excellent agreement with TCOE and $\text{GinAI}^\dagger$  across the entire spectrum, even for small $|\lambda|$, where their density of states show discrepancies, as seen in Fig.~\ref{fig:DOS_STM_COE}(a). The three distributions naturally deviate from that of the GinUE.

The agreement of the distributions of the L-QSM and TCOE with $\text{GinAI}^\dagger$, rather than GinUE, stems from the underlying COE symmetry of the closed system. Although the truncated propagator $\widetilde{U} = \hat{U} \hat{\Pi}$ in Eq.~(\ref{open_p}) is not itself transpose symmetric, its nonzero eigenvalues coincide with those of the reduced operator $\hat{\Pi} \hat{U} \hat{\Pi}$ acting on the surviving subspace. This reduced operator is complex symmetric, placing the local spectral correlations of the resonances in the AI$^\dagger$ universality class, as further explained in Appendix~\ref{subsec:leak_transpose_symmetry}.

\subsubsection{Ratio of consecutive eigenvalue spacings}
\label{sec:Rt_ls}

The ratio of consecutive eigenvalue spacings provides an alternative measure of short-range spectral correlations. Unlike spacing distributions, it does not require unfolding, which is useful in systems with nonuniform or difficult-to-model spectral densities. The complex ratio is defined as~\cite{Sa2020}
\begin{equation}
    z_{\mu} = \frac{\lambda_\mu^{NN} - \lambda_\mu}{\lambda_\mu^{NNN} - \lambda_\mu},
\end{equation}
with $\lambda_\mu^{NN}$ ($\lambda_\mu^{NNN}$) representing the nearest (next-nearest) neighbor of the eigenvalue $\lambda_\mu$ in the complex plane.

The ratio is typically expressed in the polar form $z_\mu = r_\mu e^{i\theta_\mu}$. Instead of studying the joint distributions $\mathrm{p}(r,\theta)$, we focus on the marginal distributions, that is, the probability distributions obtained by integrating either $r$ or $\theta$. In particular, $\mathrm{p}(r)$ characterizes the typical separation between neighboring eigenvalues and quantifies the strength of short-range level repulsion, while $\mathrm{p}(\theta)$  reveals geometric correlations between nearby eigenvalues and exposes angular anisotropies. 

The radial distribution $\mathrm{p}(r)$  is shown in Fig.~\ref{fig:lvlcorr_STM}(b) and the angular distribution $\mathrm{p}(\theta)$ in Fig.~\ref{fig:lvlcorr_STM}(c). 
All four cases, L-QSM (blue), TCOE (red), $\text{GinAI}^{\dagger}$ (orange), and  GinUE (black), exhibit strong level repulsion and similar $\mathrm{p}(r)$ for $r \to 0$. However, as $r$ increases, the distributions for L-QSM, TCOE, and $\text{GinAI}^{\dagger}$ progressively deviate from the GinUE result. A similar trend is observed in the angular distribution, where $\mathrm{p}(\theta)$ for L-QSM, TCOE, and $\text{GinAI}^{\dagger}$ departs from the GinUE curve, with the differences becoming most pronounced for angles close to $0$ and $\pi$.

The deviation of $\mathrm{p}(\theta)$ for L-QSM, TCOE, $\text{GinAI}^{\dagger}$ from the GinUE result suggests an enhanced probability of forming nearly collinear triplets of eigenvalues. A collinear triplet refers to three distinct eigenvalues whose positions in the complex plane lie approximately along the same straight line. In this situation, the spacing vectors from a given eigenvalue to its nearest and second-nearest neighbors have nearly the same direction (or the opposite direction), so the angle between them is close to $\theta \approx 0$ or $\theta \approx \pi$. The resulting excess probability near these angles reflects residual angular correlations in the spectrum, in contrast to the locally isotropic statistics of GinUE.

\begin{figure*}
    \centering
    \includegraphics[scale=.5]{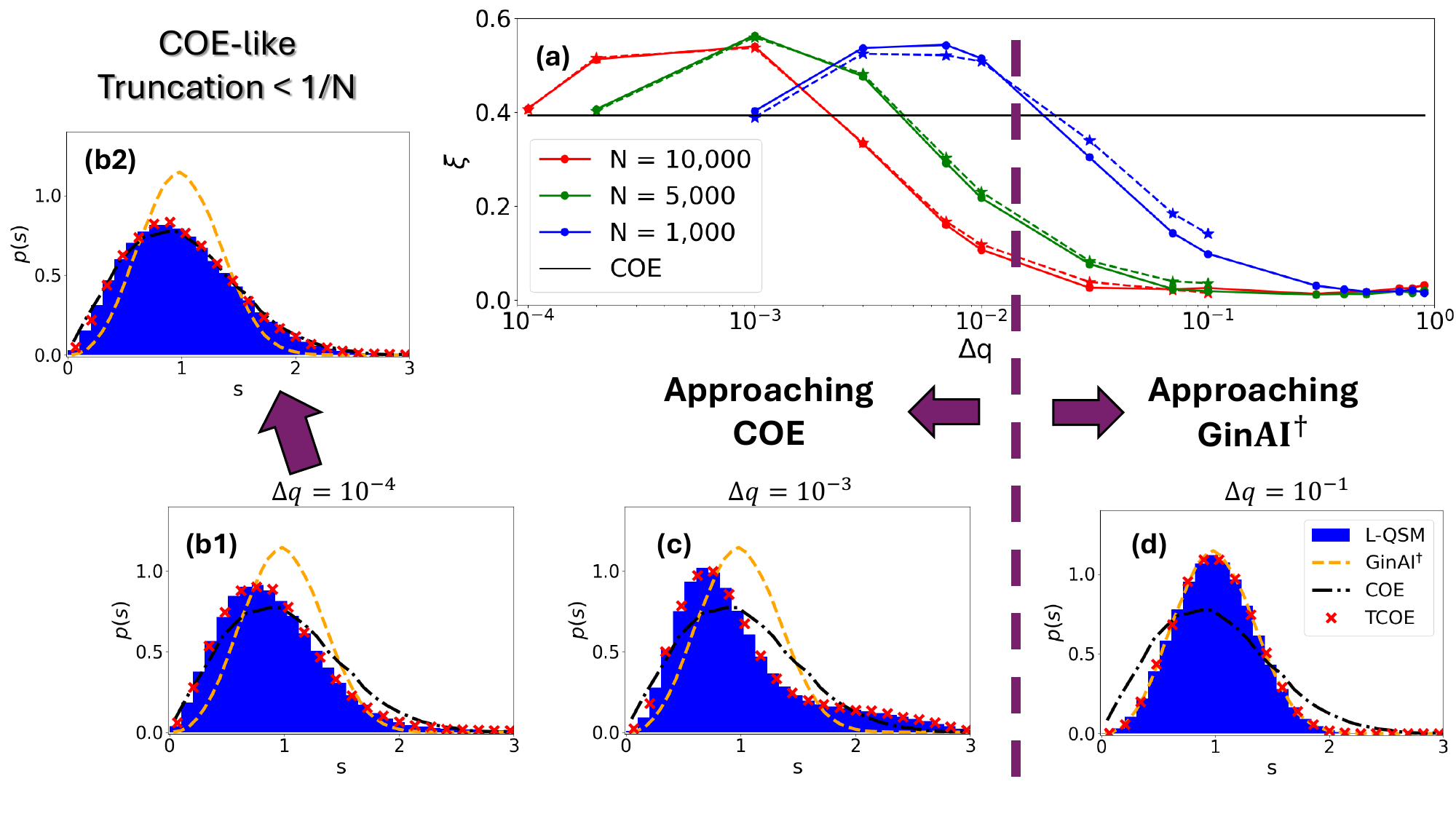}
    \caption{Short-range spectral correlations dependence on the truncation size. (a) Level-spacing indicator $\xi$ [Eq.~(\ref{Eq:LSI})], quantifying the deviation of $p(s)$ from the $\text{GinAI}^\dagger$ as a function of the leak size $\Delta q$ for different matrix dimensions $N$. Solid lines correspond to the TCOE and dashed lines to the L-QSM. The horizontal black line shows the value of $\xi$ for the COE. The nearest-neighbor spacing distributions in panels (b1), (b2), (c), and (d) use $N=10^4$. (b1) Truncation of a single column, $n_{L}=N \Delta q = 1$; (b2) partial truncation of one column; (c) truncation of $n_L=10$ columns; (d) truncation of $n_L=10^{2}$ columns, for which the L-QSM and TCOE distributions already agree with the $\text{GinAI}^\dagger$ prediction. All statistics are computed with at least $10^5$ eigenvalues.
    }
    \label{fig:lvlcorr_COE}
\end{figure*}

\section{Spectral correlations vs leak size}
\label{Sec:COEcrossover}

We now investigate how the short-range spectral correlations depend on the leak size. To this end, we introduce a level-spacing indicator that quantifies the deviation of the nearest-neighbor spacing distribution $\mathrm{p}(s)$ from the $\text{GinAI}^\dagger$ distribution,
\begin{equation}
   \xi = \frac{\sum_i |\mathrm{p}(s_i) - \mathrm{p}_{\text{AI}^\dagger}(s_i)|  }{\sum_i \mathrm{p}_{\text{AI}^\dagger}(s_i)} .
   \label{Eq:LSI}
\end{equation}
Figure~\ref{fig:lvlcorr_COE}(a) shows the indicator $\xi$ as a function of the leak size $\Delta q$ for different matrix dimensions $N$. For reference, the closed-system value (COE) is indicated by a black horizontal line.

For $N=10^3$, Fig.~\ref{fig:lvlcorr_COE}(a) shows that the agreement of $\mathrm{p}(s)$ for the L-QSM and TCOE with $\text{GinAI}^\dagger$ (small $\xi$) becomes evident for $\Delta q \geq 10^{-1}$, while for $N=10^4$, the agreement is already very good for $\Delta q \geq 10^{-2}$. This dependence  on the matrix size suggests that $\Delta q$ is not the most appropriate parameter for comparison;  instead, the relevant quantity is the number $n_L = N\Delta q$ of columns whose elements are set to zero.

\begin{figure}[t]
    \centering
    \includegraphics[scale=.33]{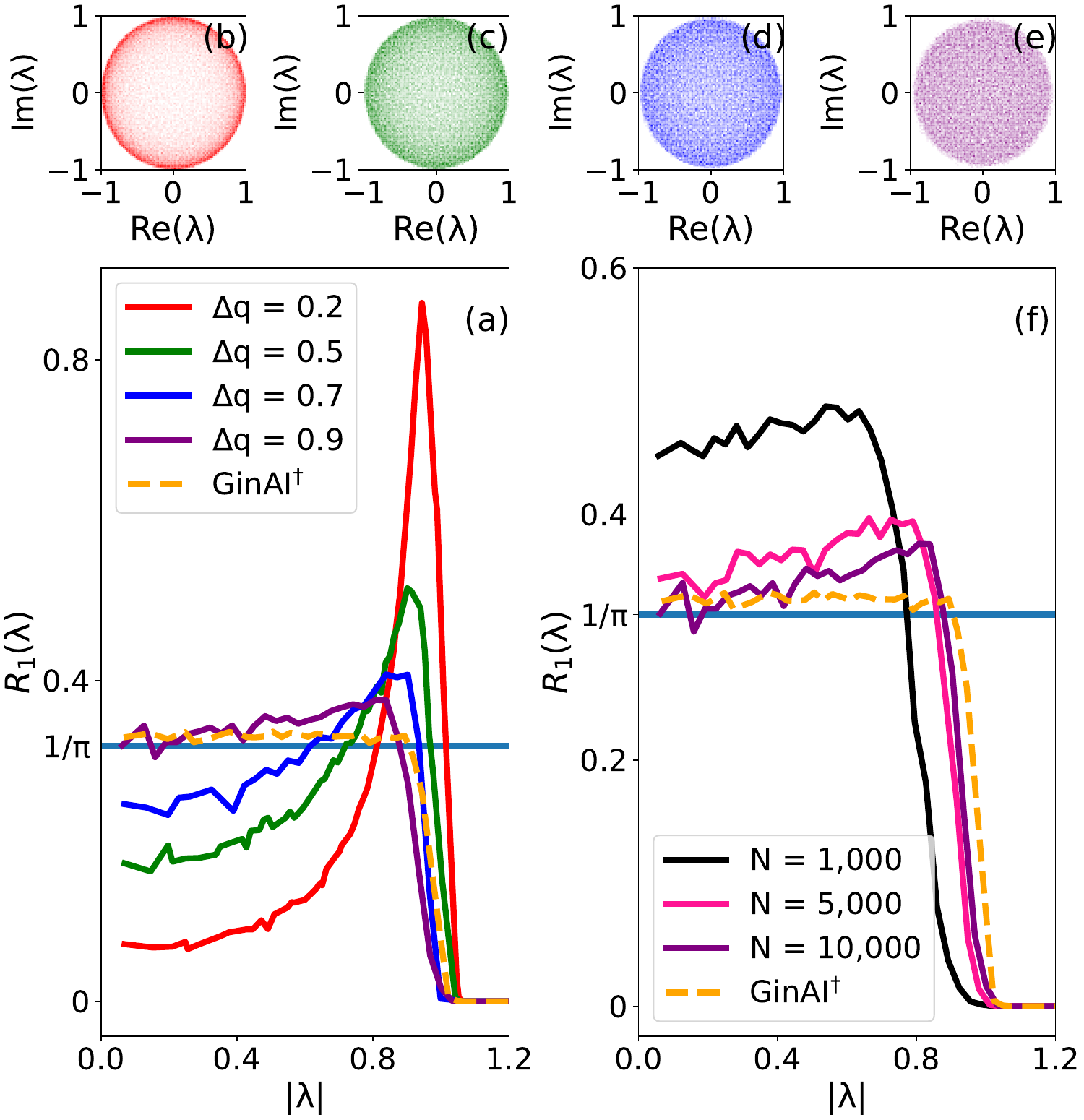}
    \caption{Convergence of the density of states $R_1(\lambda)$ of the TCOE toward the Ginibre circular law. (a) Averaged $R_1(\lambda)$ for TCOE with slit sizes $\Delta q = 0.2, 0.5, 0.7, 0.9$ (red, green, blue, and purple curve, respectively) compared with the numerical result for the $\text{GinAI}^\dagger$  (orange dashed line); $N = 10^{4}$.  (b)-(e) Distribution of the TCOE normalized eigenvalues in the complex plane for (b) $\Delta q = 0.2$, (c) $\Delta q = 0.5$, (d) $\Delta q = 0.7$, and (e) $\Delta q = 0.9$, showing progressive homogenization of the spectrum as the leakage increases. (f) Averaged $R_1(\lambda)$ for TCOE with slit size $\Delta q = 0.9$ and different dimensions  compared with the GinAI$^\dagger$ numerical result for $N = 10^{4}$ (orange dashed line).  Multiple ensemble realizations were considered to ensure at least $10^5$ eigenvalues for each dataset.
    }
    \label{fig:DOS_COE}
\end{figure}

Expressed in terms of $n_L$, the value of $\xi$ is comparable across different values of $N$. The results indicate that once $n_L \gtrsim 10^2$,  the short-range correlations of L-QSM and TCOE agree very well with $\text{GinAI}^{\dagger}$ for any matrix size $N \gtrsim 10^3$. This is illustrated in Fig.~\ref{fig:lvlcorr_COE}(d), where $N=10^4$ and $n_L=100$.

For $n_L<10^2$, the level spacing distribution of L-QSM and TCOE deviates from $\text{GinAI}^{\dagger}$ and approach the COE, as illustrated in Fig.~\ref{fig:lvlcorr_COE}(c). As the number of truncated columns decreases, the dwell time increases, allowing correlations associated with the underlying closed system to develop.

However, independently of $N \gtrsim 10^3$, there is no close agreement with COE even when only a single column is removed, as seen Fig.~\ref{fig:lvlcorr_COE}(b1). 
Even this minimal truncation produces a sizable non-unitary perturbation on the scale of local spacings. To achieve the COE limit, one needs to set to zero only a portion of a single column, as shown in Fig.~\ref{fig:lvlcorr_COE}(b2). A smooth convergence to the closed limit emerges once the effective opening strength is reduced below one full channel.

\section{Approach to the  circular law}
\label{Sec:LargeLeakage}

As demonstrated in the preceding sections, the TCOE provides an effective random-matrix benchmark for the L-QSM, accurately capturing both its density of states and spectral correlations. Furthermore, for $n_L>10^2$, the short-range spectral correlation of both agree with the GinAI$^\dagger$ result.

The question now addressed is whether increasing the leakage size can push the global spectral properties toward the Ginibre prediction. This question is motivated by the analytical results in Ref.~\cite{Zyczkowski2000}, where it was shown that the density of states of non-Hermitian matrices of size $M$, constructed as square submatrices of unitary random matrices of size $N$, converge to those of Ginibre ensembles as the ratio $M/N$ becomes sufficiently small.

Because the number of surviving resonances decreases dramatically at large leakage, the direct analysis of the L-QSM becomes computationally prohibitive even after renormalizing the spectrum by $\lambda_{\text{max}} = \max{(\{\lambda_\mu\})}$. For instance, only $0.01\%$ of resonances remain at $\Delta q = 0.9$. In this strongly open regime, the TCOE provides a tractable surrogate for the L-QSM.


Figure~\ref{fig:DOS_COE}(a) shows how the smoothed density of states of the TCOE converges to the Ginibre circular law as the slit width $\Delta q$ increases. For moderate leakage ($\Delta q = 0.2$, red line), long-lived resonances persist and accumulate near the unit circle [Fig.~\ref{fig:DOS_COE}(b)], so $R_1(\lambda)$ strongly deviates from the uniform $1/\pi$ profile. As the leak grows ($\Delta q = 0.5$, $0.7$, and $0.9$; green, blue, and purple curves), the distribution undergoes a qualitative transition: The accumulation near $|\lambda|\approx 1$ gradually disappears [cf. Figs.~\ref{fig:DOS_COE}(b)-(e)], and the distribution becomes increasingly uniform, approaching the Ginibre limit. This trend is consistent with earlier predictions from non-unitary random matrix theory in~\cite{Zyczkowski2000,Fyodorov1997,Fyodorov1999}. The convergence is further enhanced with increasing matrix dimension, as illustrated in Fig.~\ref{fig:DOS_COE}(f) for $\Delta q = 0.9$.  

\section{Conclusions}
\label{Sec:Conclusions}

The paper investigates the spectral properties of strongly chaotic Floquet systems with localized leaks, focusing on the leaky quantum standard map (L-QSM) and comparing it with the truncated circular orthogonal ensemble (TCOE) as a random-matrix benchmark. The main goal is to understand how localized escape affects the local spectral correlations and the global distribution of the eigenvalues.

The study shows that the short-range spectral correlations in the bulk of the spectrum of the L-QSM and the TCOE are equivalent and follow the universal statistics of the non-Hermitian symmetry class AI$^\dagger$ (denoted by GinAI$^\dagger$), rather than the statistics of the unconstrained Ginibre ensemble (GinUE). This reflects the presence of transpose symmetry inherited from the time-reversal symmetry of the closed system.

Interestingly, the agreement with GinAI$^\dagger$ statistics is not controlled by the truncation ratio of the L-QSM and TCOE matrices, but rather by the absolute number of removed columns. Consequently, as the matrices increase, smaller leaks are sufficient to drive the spectral correlations toward those of the GinAI${}^{\dagger}$ ensemble. Furthermore, the COE limit is achieved only when the truncation is smaller than one full column. This analysis provides new insight into the interplay between residual unitary structure and the emergent non-Hermitian spectral statistics induced by truncation.

Although local spectral correlations quickly converge to the AI$^\dagger$ universality class, the approach toward the Ginibre circular law requires a large leakage. This reveals a clear distinction between the dependence of the local and global spectral properties of the Floquet systems on the size of the localized leak. 

Our results clarify the role of symmetry constraints in the spectral properties of open Floquet systems and has direct relevance for 
experimental platforms where loss is spatially localized. Interesting future directions include exploring multiple leaks and time-dependent openness.


\begin{acknowledgments}
This research was supported by the
Research Corporation Cottrell SEED award CS-SEED-2025-003. EMS and LFS are very grateful to Gernot Akemann for valuable discussions on complex-symmetric Ginibre random matrices.
\end{acknowledgments}


\section{Code and Data Availability}
\label{Sec:Data}

All the codes and data used to generate the figures in the manuscript  analysis and dynamical simulations can be found at \href{https://github.com/edsonsignor/QSM-leakage}{https://github.com/edsonsignor/QSM-leakage}.


\appendix

\section{Classical and quantum standard map}
\label{App:QSM}

\begin{figure}[h]
    \centering
    \includegraphics[scale=.4]{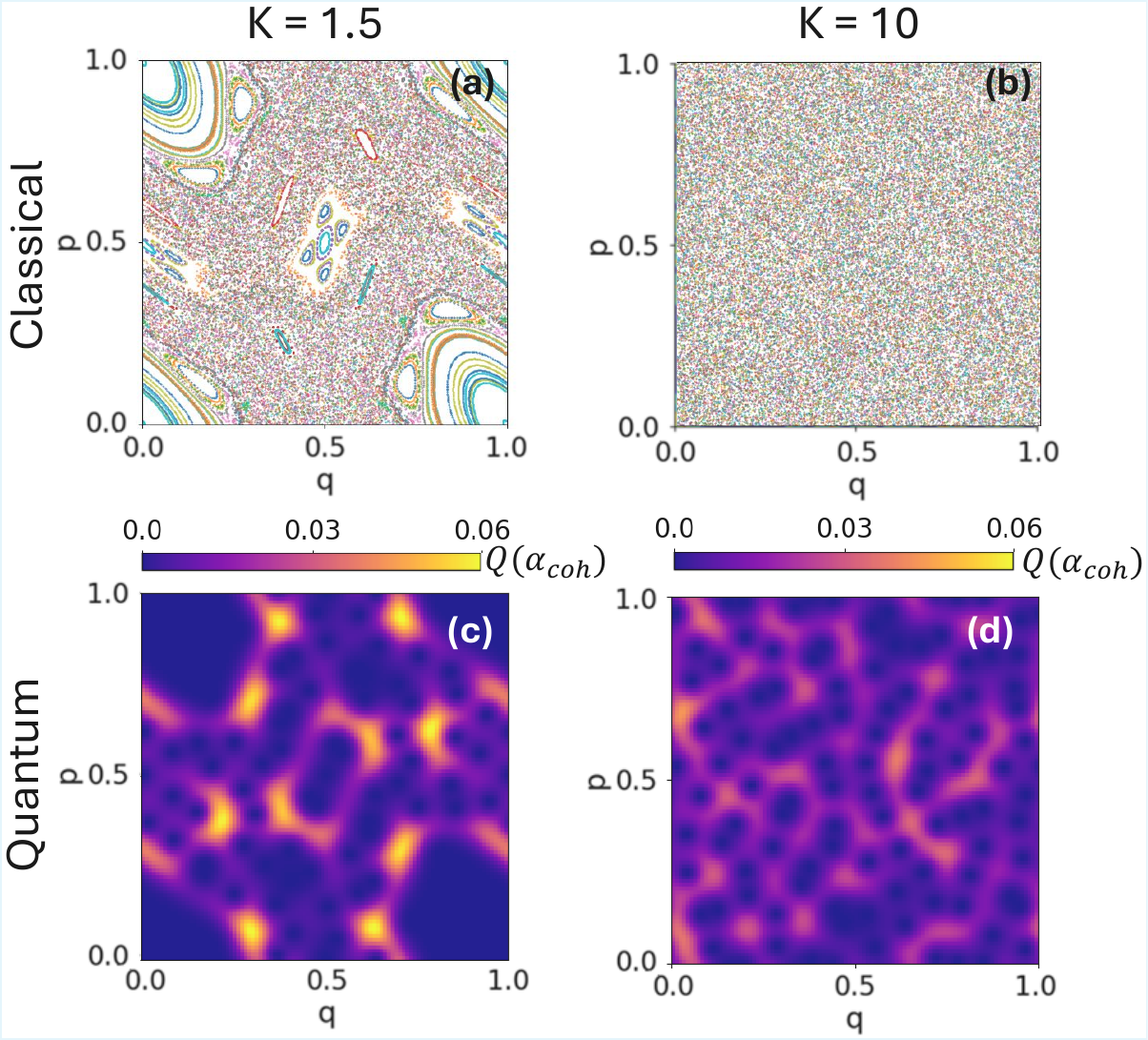}
    \caption{Classical-quantum correspondence of the closed standard map for (a),(c) mixed phase-space regime at $K=1.5$ and (b),(d) fully chaotic regime at $K=10$. 
    (a)-(b) Classical phase-space landscapes and (c)-(d) Husimi Q-functions of selected eigenstates. 
    }
    \label{fig:ClosedSM}
\end{figure}

Figure~\ref{fig:ClosedSM} shows the results for the standard map with two values of the kick strength $K$. The phase space of the classical standard map, following Eq.~(\ref{eq:SM}), are shown in Figs.~\ref{fig:ClosedSM}(a)-(b). For small $K$ the phase space contains regular islands embedded in a chaotic sea, as illustrated in Fig.~\ref{fig:ClosedSM}(a) for $K = 1.5$. For sufficiently large $K$ the motion becomes globally chaotic, as shown in Fig.~\ref{fig:ClosedSM}(b) for the value $K=10$ used throughout this work.

{\bf Quantization on the torus:} As discussed in the main text, quantization on the torus leads to a Hilbert space of dimension
\[
N=\frac{1}{2\pi\hbar}.
\]
The toroidal phase space leads to periodic boundary conditions in both position and momentum and the variables become discrete
\[
q_n = \frac{n+\varphi}{N}, \quad \quad p_m = \frac{m}{N}, \quad \quad n,m = 1, \dots, N,
\]
where $\varphi$ specifies the boundary-condition phase associated with torus quantization.

The system evolves under the Floquet operator of the quantum standard map, which for a unit period takes the form
\begin{equation}
\hat{U} = \hat{U}_{\mathrm{kick}}\, \hat{U}_{\mathrm{free}}
= \exp\!\left[\frac{iK}{4\pi^2\hbar}\cos(2\pi \hat{q})\right]
\exp\!\left[-\frac{i}{\hbar}\frac{\hat{p}^2}{2}\right],
\label{eq:F_operator}
\end{equation}
where $\hat{U}_{\mathrm{kick}}$ represents the instantaneous kick and $\hat{U}_{\mathrm{free}}$ the free evolution between two kicks ~\cite{Lakshminarayan2018}.
On the discrete torus, in the position representation, 
these operators are given by
\begin{align}
\langle q_n|\hat{U}_{\mathrm{kick}}|q_{n'}\rangle &= \exp\!\left\{i\frac{K N}{2\pi}\cos\!\left[\frac{2\pi}{N} ( n+\varphi)\right]\right\} \delta_{n,n'} \\
\langle q_n|\hat{U}_{\mathrm{free}}|q_{n'}\rangle
&= \sum_m \langle q_n|p_m\rangle \exp\!\left[-i\frac{\pi m^2}{N}\right] \langle p_m|q_{n'}\rangle. 
\end{align}
Using the discrete Fourier overlap,
\begin{equation}
\langle q_n | p_m \rangle
= \frac{1}{\sqrt{N}}
\exp \left[
   \frac{2\pi i}{N} m \,(n + \varphi)
\right],
\label{eq:FourierOverlap}
\end{equation}
and the quadratic Gaussian sum,
\begin{align}
\frac{1}{N}\sum_{m=0}^{N-1}
e^{
    -\,i\frac{\pi}{N}m^2
    +\,i\frac{2\pi}{N}m(n-n')
}
= \sqrt{\frac{i}{N}}\,
e^{
    i\,\frac{\pi}{N}(n - n')^2
},
\label{eq:GaussSum}
\end{align}
we obtain the Eq.~(\ref{eq:QSM_closed}) of the main text:
\begin{equation}
U_{n,n'} \!= 
\sqrt{\frac{i}{N}} 
\exp \!\left\{ \!
    \frac{i\pi}{N}(n-n')^2 
    + \frac{iNK}{2\pi} 
    \cos \! \left[ \frac{2\pi}{N}(n+\varphi) \right] \!\right\} .
\end{equation}
 
{\bf Quantum-classical correspondence:} The quantum-classical correspondence becomes evident when the eigenstates of the Floquet operator are represented through their Husimi $Q$-functions.  
Figure~\ref{fig:ClosedSM} compare the quantum and classical results.
For $K = 1.5$ [Fig.~\ref{fig:ClosedSM}(c)], the Husimi distribution of a typical eigenstate is concentrated in the regular islands, reflecting the mixed phase-space structure of the classical dynamics. For $K = 10$ [Fig.~\ref{fig:ClosedSM}(d)], the Husimi function becomes delocalized over the entire chaotic sea, consistent with the ergodic character of the underlying classical motion~\cite{Saraceno1990}.


\section{Spectral filtering}
\label{sec:bulk_selection}

When comparing the spectrum of a physical system with random-matrix ensembles, it is important to exclude system-specific features that obscure universal behavior. In the L-QSM, such nonuniversal contributions arise from fast-decaying states. The structures of these states are correlated with the location of the leak. Rather than spreading ergodically over phase space, they stem from non-ergodic Floquet states of the closed system that concentrate near the region where the leak is introduced. These short-lived resonances produce the dense cluster of quasienergies near the origin of the complex plane in Fig.~\ref{fig:Ldistri}(c). They have been extensively studied in previous works~\cite{Schomerus2004,Hall2023,Pedrosa2009,Nonnenmacher2007}, but are excluded here, as our analysis focuses on the bulk of the spectrum and its universal properties.

To determine the exclusion threshold, we use the smallest absolute values of the Floquet eigenvalues $|\lambda|$ obtained from the TCOE with the same leak size $\Delta q$ as the L-QSM. Unlike the physical quantum map [Fig.~\ref{fig:OpenSM}(d)], the dwell times of the TCOE are independent of the leak position by construction. Hence, the minimal $|\lambda|$ values of the TCOE provide a natural benchmark for decay rates originating purely from stochastic openness, free of dynamical correlations.

Following this criterion, we discard eigenvalues with $|\lambda| < 10^{-1}$ 
(corresponding to dwell times $\tau < 0.22$), retaining only the subset of resonances expected to display universal chaotic statistics. This filtering process significantly increases the computational cost of the spectral analysis. For a leak size of $\Delta q = 0.2$, roughly half of the spectrum must be excluded, and the fraction grows rapidly as the leak becomes larger. To compensate for the reduced number of usable eigenvalues, we construct an ensemble of L-QSM spectra by varying the kick strength within the range $9.99 < K < 10.01$, using increments that ensure a total data set of at least $10^5$ eigenvalues. 

Finally, we further normalize each spectrum $\{\lambda_\mu \}$ of the L-QSM and of the TCOE by its maximal eigenvalue $\lambda_{\text{max}} = \max{(\{\lambda_\mu\})}$. This rescaling removes differences in the overall decay scale while preserving the spectral structure.


\section{Unfolding of complex spectra}
\label{App:Unfolding}

We need to unfold the spectrum  to separate system-specific spectral features from universal fluctuations. For a spectrum with density of states 
$R_1(\lambda)$, this decomposition reads
\begin{equation}
    R_1(\lambda) =  \bar{R}_1(\lambda) + \delta R_1(\lambda),
\end{equation}
where $\bar{R}_1(\lambda)$ captures the smooth, nonuniversal part of the density of states, and $\delta R_1(\lambda)$ encodes the universal correlations of interest.

Several approaches have been proposed for estimating $\bar{R}_{1}(\lambda)$ \cite{Akemann2019,Mudute-Ndumbe2020, Jisha2024}. In the case of locally isotropic spectra, $\bar{R}_{1}$ can be approximated by a sum of Gaussians~\cite{Akemann2019}. If the eigenvalue distribution depends only on the radius, not on the angle, one can fit the radial profile or use a theoretical closed form~\cite{Mudute-Ndumbe2020,Jisha2024}.

Here, we employ the well-established local rescaling method described in~\cite{Haake1992, Grobe1988, HaakeBook}, which is accurate and computationally efficient. The  local mean density is estimated as
\begin{equation}
    \bar{R}_1(\lambda_\mu) = \frac{k}{\pi d_{k,\mu}^2},
\end{equation}
where $k$ is an integer chosen in the range $1 \ll k \ll N$ and $d_{k,\mu}$ is the distance from $\lambda_\mu$ to its $k$-th nearest neighbor eigenvalue in the complex plane. In our calculations, we choose $k=30$. 
This procedure amounts to computing the area of a disk of radius $d_{k,\mu}$ centered at $\lambda_\mu$ and then obtaining the unfolded spacings 
through the local rescaling \footnote{This method is equivalent to that used for Hermitian matrices described in~\cite{Gubin2012}, where each subset of $k$ consecutive level spacings, $s_{i+1}, \ldots s_{i+k}$, are divided by the local mean spacing given by $(E_{i+k+1}-E_{i+1})/k$. We usually choose $1\ll k\lesssim 10$.},
\begin{equation}
    \tilde{s}_\mu = s_\mu \sqrt{\bar{R}_1(\lambda_\mu)}.
\end{equation}
Boundary effects vanish in 
the limit $N \to \infty$ with $k/N \to 0$.

The unfolded spacing distribution 
$\mathrm{p}(\tilde{s})$ is normalized as
\begin{equation}
    \int_0^\infty \mathrm{p}(\tilde{s})\, d\tilde{s} = 
    \int_0^\infty \tilde{s}\, \mathrm{p}(\tilde{s})\, d\tilde{s} = 1.
\end{equation}


\section{Symmetry of the truncated propagator}
\label{subsec:leak_transpose_symmetry}

We explain why the comparison with the non-Hermitian symmetry class
$\mathrm{AI}^\dagger$ is appropriate for the nonzero resonance spectrum
of the truncated propagators of the L-QSM and TCOE. 

Although the open-system propagator $\hat U\hat\Pi$ defined in
Eq.~(\ref{open_p}) is generally not complex symmetric, its nonzero
eigenvalues coincide with those of the reduced operator $\hat\Pi\hat U\hat\Pi$ acting on the surviving subspace. The latter operator is complex symmetric when the closed system belongs to the COE
symmetry class.

To see this explicitly, reorder the basis such that the surviving states come first and the leaked states last. Then
\begin{equation}
\hat{\Pi}=
\begin{pmatrix}
I_{N-n_{L}} & 0\\
0 & 0_{n_{L}}
\end{pmatrix},
\qquad
\hat{U}=
\begin{pmatrix}
A & C\\
B & D
\end{pmatrix},
\label{eq:block_U_Pi}
\end{equation}
where $n_{L}$ is the number of removed columns. It follows that
\begin{equation}
\hat{U}\hat{\Pi}=
\begin{pmatrix}
A & 0\\
B & 0
\end{pmatrix}.
\label{eq:UPi_block}
\end{equation}
The characteristic polynomial factorizes as
\begin{align}
\det(\lambda I_N-\hat{U}\hat{\Pi})
&=
\det
\begin{pmatrix}
\lambda I_{N-n_{L}}-A & 0\\
-B & \lambda I_{n_{L}}
\end{pmatrix}
\nonumber\\
&=
\lambda^{n_{L}} \det(\lambda I_{N-n_{L}}-A).
\label{eq:charpoly_factorization}
\end{align}
Therefore, $\hat{U}\hat{\Pi}$ has $n_{L}$ zero eigenvalues, and its nonzero eigenvalues are precisely the eigenvalues of the block $A$. On the other hand,
\begin{equation}
\hat{\Pi} \hat{U} \hat{\Pi}=
\begin{pmatrix}
A & 0\\
0 & 0
\end{pmatrix},
\label{eq:PiUPi_block}
\end{equation}
so the nonzero spectrum of $\hat{U} \hat{\Pi}$ is exactly the spectrum of the reduced operator $\hat{\Pi} \hat{U} \hat{\Pi}$ on the surviving subspace.

If the closed Floquet operator belong to the COE symmetry class, then there is a symmetry-adapted basis in which 
\[
\hat{U}^T=\hat{U}.
\]
Since the projection operator is diagonal in the chosen basis,
$\hat\Pi=\hat\Pi^T$, and it follows that
\begin{equation}
(\hat{\Pi}\hat{U}\hat{\Pi})^T = \hat{\Pi} \hat{U}^T \hat{\Pi} = \hat{\Pi}\hat{U}\hat{\Pi},
\label{eq:QUQ_symmetric}
\end{equation}
so the reduced operator is complex symmetric. 

In our spectral analysis, the exact zero eigenvalues  generated by the truncation are excluded. Therefore, although the truncated propagator $\hat U\hat\Pi$ is not itself symmetric, the nonzero resonances are governed by a complex
symmetric non-Hermitian operator. This places their local spectral
statistics in the universality class $\mathrm{AI}^\dagger$.


\bibliography{biblio2025}

\end{document}